\documentclass[journal]{IEEEtran}
\usepackage{amsmath,amssymb}
\usepackage[dvips]{graphicx}
\usepackage{amsfonts}
\usepackage[mathscr]{eucal}
\usepackage{latexsym}
\usepackage{amsthm}
\usepackage{exscale}
\usepackage[mathscr]{eucal}
\usepackage{bm}
\usepackage[dvipsnames]{color}
\usepackage{cases}
\usepackage{epsfig}
\usepackage[center,small]{caption}
\usepackage[ruled,linesnumbered]{algorithm2e}
\usepackage[verbose,nospace,sort]{cite}
\usepackage{tabularx}
\usepackage{multirow}
\usepackage{balance}
\usepackage{url}
\scrollmode

\SetCommentSty{mycommfont}

\newtheorem{definition}{Definition}
\newtheorem{lemma}{Lemma}

\begin{document}

\title{Single Threshold Packet Scheduling Policy for AoI Minimization in Resource-Constrained Network
}
\author{\IEEEauthorblockN{Yonghao~Ji
			and Xiaoli~Xu  \\
		{School of Information Science and Engineering, Southeast University, Nanjing, China \\
			Email: \{220210839, xiaolixu\}@seu.edu.cn}}}

\maketitle

\begin{abstract}
This paper investigates the tradeoff between the average age of information (AoI) and the transmission cost for networks with stochastic packet arrival and random erasure channel. Specifically, we model the resource-constrained AoI minimization problem as a constrained Markov decision process (CMDP) and propose a low-complexity single threshold packet scheduling policy for it. The key advantage of the proposed policy is its tractability and convenience for implementation. The AoI distribution and long-term average transmission cost of the proposed policy are derived as closed-form functions of the selected threshold. Furthermore, we show that the proposed policy reduces to the optimal policies under special settings and achieves close-to-optimal performance under general settings.  
\end{abstract}
\section{Introduction}
Data freshness is of paramount importance in real-time communication applications, such as automatic driving, remote sensing and control. Age of information (AoI) is a widely used  metric to quantize the information freshness at the receiver \cite{ref1}. The pioneer works \cite{ref2}\cite{ref3}\cite{ref4} mainly considered the average AoI performance under different queuing policies, e.g., the first-come-first-serve (FCFS) and the last-generated-first-serve (LGFS) queues. It was pointed out in \cite{Bedewy2016} that the preemptive LGFS (pLGFS) achieves the minimum average AoI in point-to-point networks with stochastic packet arrival. The analysis of AoI under different queuing policies was extended to the discrete-time networks in \cite{ref4}. To capture the effect of AoI on the real applications, general age penalty functions and the corresponding optimal policies were investigated in \cite{Sun2017}. A typical example of the age penalty function is the age violation probability raised in \cite{ref7}. Evaluating the general age penalty function usually requires not only the average AoI, but also the complete AoI distribution. Reference \cite{ref5} and \cite{ref6} analyzed the AoI distribution under various packet arrival models and queuing policies, for both continuous and discrete queues.

The tradeoff between the AoI and other network performance metrics has also attracted a lot of research interest, such as the tradeoff between AoI and energy consumption \cite{ref15}, and the tradeoff between AoI and packet delivery ratio\cite{XU2021}. Besides, optimization of AoI under resource constraint is of particular interest for wireless sensor networks with limited power and/or spectrum. The authors in \cite{ref14} considered a resource-constrained network with ``generate-at-will" packet arrival model and random erasure channel. They proved that the threshold structure packet scheduling policies can achieve the optimal average AoI under both the automatic repeat request (ARQ) and hybrid ARQ transmission protocols. Reference \cite{ref9} and \cite{ref10} considered the AoI minimization in two-hop relaying networks, with total resource constraint applied on the source and relay nodes. A double threshold method was shown to achieve the near optimal AoI performance for the two-hop network when the source packet is generated at will \cite{ref9}.

When the packets can be generated at will by the transmitter, the design of the packet scheduling policy only needs to consider the receiving status of the previous transmissions. However, the practical data arrival may be controlled by independent component of the circuit and subject to certain constraint, e.g., the sensing power limitation. In this paper, we extend the existing works by considering the resource-constrained network with stochastic packet arrival. We formulate the optimization of AoI under resource constraint as a constrained Markov decision process (CMDP), and propose a lower bound of the AoI as a function of the long-term average transmission cost. Furthermore, we propose a low-complexity single threshold packet scheduling policy and derive the AoI distribution, average AoI and average transmission cost as the closed-form functions of a single threshold. With the closed-form performance characterizations, the optimal threshold can be easily determined for any given constraint and the proposed policy can be implemented in a very efficient manner. We show that when the resource constraint is relaxed or the packet arrival rate is high enough to mimic the ``generate-at-will" model, the proposed single threshold policy reduces to the corresponding optimal schemes presented in \cite{Bedewy2016} and \cite{ref9}, respectively. Finally, extensive simulations are conducted to verify the analytical results and compare the performance of the proposed policy with exiting benchmarks.

\section{System Model}

\begin{figure}[htbp]
	\centering
	\scalebox{0.65}{\includegraphics{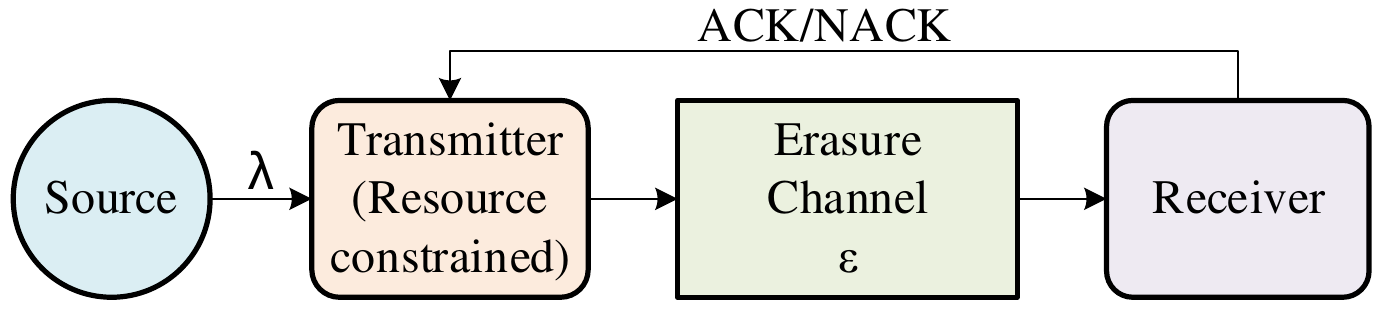}}
	\caption{System model.}
    \label{F:system}
\end{figure}

As shown in Fig.~\ref{F:system}, we consider a slotted point-to-point communication system, where the updating packets are generated from the source at the beginning of each time slot. The arrival of the packets to the transmitter follows Bernoulli process with arrival rate $\lambda$ and the arriving time cost is neglected. Each packet arrived at the transmitter is stored in a unit-size buffer, and sent by the transmitter to the receiver through an erasure channel with erasure probability $\varepsilon$.
If the transmission is successful, the packet will reach the receiver side at the end of the same time slot and an instantaneous error-free feedback will be received by the transmitter at the same time. The packet is removed from the buffer if it is successfully delivered or preempted by the newly arrived packet.

Note that each packet transmission, regardless whether it is successful, will cost certain amount of transmission power and occupy the channel spectrum and time. Hence, it is desired to minimize the transmission cost, while maintain the acceptable data freshness at the receiver.

\subsection{Data Freshness}
At the transmitter side, the data freshness is only affected by the packet arrival rate $\lambda$. We assume that the transmitter data freshness at the $i$th time slot is denoted by $\Delta_t(i)$ and measured at the beginning of each time slot. If there is a packet arrival at the $i$th time slot, we have $\Delta_t(i)=0$. Otherwise, we have
\begin{equation}
	\Delta_t(i)=i-G(i) \label{eq:AoIt}
\end{equation}
where $G(i)$ is the arrival time of the latest packet observed at the $i$th time slot.

At the receiver side, the data freshness is not only affected by the packet arrival process, but also by the transmission policy and channel quality. We assume that the data freshness at the receiver is measured by the observed AoI at the end of each time slot. Denoted by $\Delta_r(i)$ the AoI observed at the $i$th time slot. Following the conventional definition of AoI, we have:
\begin{equation}
\Delta_r(i)=i-U(i)+1 \label{eq:AoIr}
\end{equation}
where $U(i)$ is the arrival slot of the latest packet successfully received by the end of $i$th slot. For consistence, we initialize $\Delta_t(0)=\Delta_r(0)=0$ and adopt the average AoI as the measurement of long-term data freshness, i.e.,
\begin{align}
\bar{\Delta}_r=\lim_{T\rightarrow\infty}\frac{1}{T}\sum_{i=1}^{T}\Delta_r(i),
\end{align}
where $T$ is the observation interval.
\subsection{Long-term Average Transmission Cost}
The transmission cost is proportional to the number of transmissions made by the transmitter.  We denote the transmitter action at the $i$th time slot by $a_i\in\{0,1\}$, where $a_i=1$ if the transmitter chooses to send the packet in the buffer and $a_i=0$ if the buffer is empty or the transmitter chooses to remain silent even though a packet is waiting in the buffer. The long-term average transmission cost is defined as
\begin{align}
\eta=\lim_{T\rightarrow\infty}\frac{1}{T}\sum_{i=1}^{T}a_i.
\end{align}

This paper aims to investigate the tradeoff between the data freshness  $\bar{\Delta}_r$ and the transmission cost $\eta$. The conventional pLGFS without resource constraint \cite{Bedewy2016} achieves the minimum AoI with the maximum transmission cost. Hence, we have
\begin{align}
\bar{\Delta}_r\geq\frac{1}{\lambda}+\frac{\varepsilon}{1-\varepsilon},\textnormal{ and }
\eta\leq\frac{\lambda}{1-(1-\lambda)\varepsilon}, \label{eq:pLGFS}
\end{align}
where the equality holds when the pLGFS transmission policy is adopted. On the other hand, when the packet is generated-at-will, which is equivalent to the proposed network model with $\lambda=1$, the optimal tradeoff between the data freshness and transmission cost has been presented in \cite{ref14}, which is reproduced as following
\begin{align}
&\bar{\Delta}_r(\delta)=\frac{(\delta(1-\varepsilon)+\varepsilon)^2+\varepsilon}{2(1-\varepsilon)(\delta(1-\varepsilon)+\varepsilon)}+\frac{1}{2},\label{eq:AoIARQ}\\
&\eta(\delta)=\frac{1}{\delta(1-\varepsilon)+\varepsilon},\label{eq:etaARQ}
\end{align}
where $\delta$ is the system design parameter and various tradeoff between $\bar{\Delta}_r$ and $\eta$ can be achieved by changing  $\delta$. This paper aims to generalize the result of  \cite{ref14} to the network with stochastic arrival.

\section{Problem Formulation and Analysis}
To investigate the tradeoff between data freshness and transmission cost, we consider the AoI minimization problem subject to the long-term transmission cost constraint, i.e.,
\begin{align}
\mathbf{P0:}\quad &\quad \min_{\{a_i\}} \bar{\Delta}_r\\
\textnormal{s.t.,}&\quad \eta\leq \eta_{\max},
\end{align}
where $\eta_{\max}$ is the maximum long-term average transmission cost. By varying the constraint $\eta_{\max}$ and solving $\mathbf{P0}$, we can obtain the trade-off between AoI and transmission cost.

\subsection{CMDP Modeling}
The problem $\mathbf{P0}$ can be modeled as a CMDP as in \cite{ref14} and \cite{ref9}. For the network model considered in this paper, the CMDP consists of the following key elements:
\begin{itemize}
\item{\emph{State}: Under the assumption of random packet arrival and erasure channel, the optimal action of the transmitter at a typical time slot is affected by the data freshness observed by the transmitter at the beginning of this time slot and that observed by the receiver at the end of the previous time slot. Hence, the state of the environment can be characterized by $s_i=(\Delta_t(i),\Delta_r(i-1))$.}
\item{\emph{Action}: At the $i$th time slot, the transmitter can choose to transmit the packet in the buffer or remain silent, corresponding to $a_i=1$ and $a_i=0$, respectively. }
\item{\emph{Reward}: Since the objective is to minimize the average AoI observed by the receiver, we define the immediate reward at the $i$th time slot as the negative of the receiver AoI, i.e., $R_i=-\Delta_r(i)$.  }
\item{\emph{Cost}: Each transmission incurs a unit cost and hence we can define the cost at the $i$th time slot, denoted by $C_i$, as the action taken, i.e., $C_i=a_i$.}
\end{itemize}

\begin{figure}[htbp]
	\centering
	\scalebox{0.5}{\includegraphics{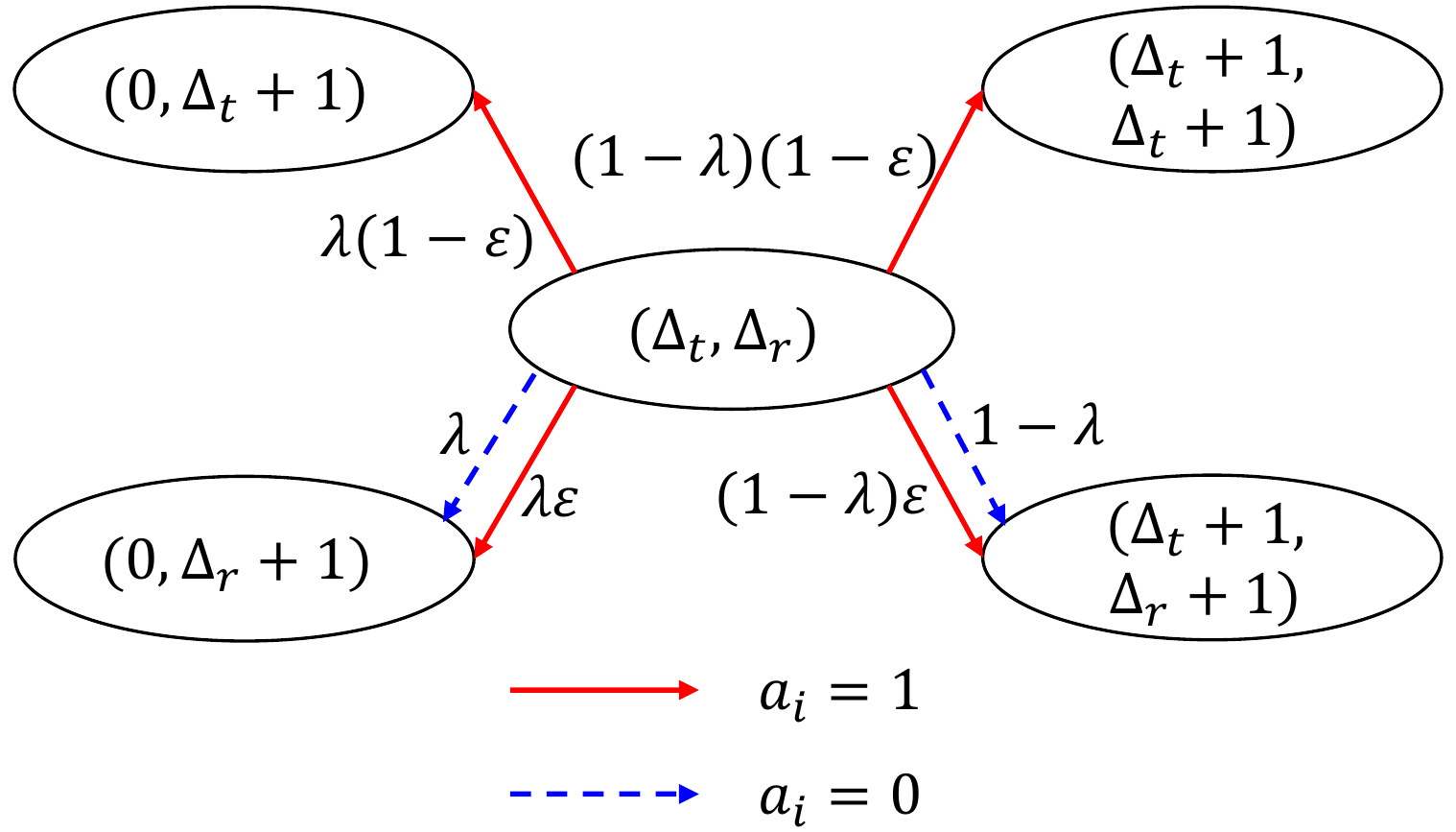}}
	\caption{The state transition diagram of the CMDP. The transition probability is labelled on the corresponding transition arrows. }
    \label{F:transition}
\end{figure}

The state transition probability is related with the random packet arrival, channel status and action taken, which is illustrated in Fig.~\ref{F:transition}.

%Finally the CMDP problem is defined as follows
%
%\emph{Problem1}:
%\begin{equation}
%\min_{\pi}J_{\pi}\triangleq\lim_{I\rightarrow \infty}\frac{1}{I}\mathbb{E}_{\pi}[\sum_{i=1}^{I}\delta_{r,i}]  \label{eq4}
%\end{equation}
%\begin{equation}
%\text{s.t.} \ \ \ \ C_{\pi}\triangleq\lim_{I\rightarrow \infty}\frac{1}{I}\mathbb{E}_{\pi}[\sum_{i=1}^{I}u_i] \leq C_{max} \label{eq5}
%\end{equation}
%
%where the expectation is respect to the policy $\pi$, $C_{max} \in (0,1]$ is the maximum value of average channel resource utilization.

In summary, solving $\mathbf{P0}$ is equivalent to find the optimal policy for the above CMDP. The classical relative value iteration algorithm (RVIA)\cite{ref17} can be used to find the optimal policy of CMDP by reducing the original infinite state space to finite size as approximation. However, as the optimal policy derived by RVIA has a multiple threshold structure \cite{ref9}, its implementation complexity is high and tractability is limited. In this paper, we derive a lower bound on the AoI under a given resource constraint and propose a low-complexity single threshold policy, which allows us to have a near optimal solution to problem $\mathbf{P0}$ in closed form, and hence investigate the tradeoff between AoI and transmission cost easily.

\subsection{Lower Bound on the AoI}
Suppose there are $N(T)$ packets successfully delivered to the receiver during the observation interval $T$. We denote by $Y_j$ the number of slots between the receiving time of the $j$th packet and the $(j+1)$th packet, and let $Y_0$ be the number of time slots before receiving the first packet. For notational convenience, we further denote by $Y_{N(T)}$ the remaining number of time slots after receiving the last packet, i.e., $Y_{N(T)}=T-\sum_{j=0}^{N(T)-1}Y_j$.

Let $S_j$ be the age of the $j$th packet when it is successfully delivered, where $S_0=0$. The receiver AoI during the time slots between the delivery of the $j$th and $(j+1)$th packet are $S_j, S_{j}+1,...,S_{j}+Y_j$, respectively. Hence, the long-term average AoI is given by
\begin{align}
\bar{\Delta}_r&=\lim_{T\rightarrow\infty}\frac{1}{T}\sum_{j=0}^{N(T)}\frac{(S_j+S_j+Y_j+1)Y_j}{2}\nonumber\\
&=\lim_{T\rightarrow\infty}\frac{1}{T}\left(\sum_{j=0}^{N(T)}S_jY_j+\frac{1}{2}\sum_{j=0}^{N(T)}Y_j^2+\frac{1}{2}\sum_{j=0}^{N(T)}Y_j\right)\nonumber\\
&\overset{(a)}{\geq}\lim_{T\rightarrow\infty}\frac{1}{T}\left(\frac{1}{2}\sum_{j=0}^{N(T)}Y_j^2+\frac{T}{2}\right)\nonumber\\
&=\frac{1}{2}\lim_{T\rightarrow\infty}\frac{1}{T}\sum_{j=0}^{N(T)}Y_j^2+\frac{1}{2}\label{eq:AoIboundder}
\end{align}
where $(a)$ follows from the fact that $\sum_{j=0}^{N(T)}S_jY_j\geq 0$ and $\sum_{j=0}^{N(T)}Y_j=T$.

Denote the expectation of $Y_j$ by $\mathbb{E}[Y_j]$, i.e., $\mathbb{E}[Y_j]\triangleq\frac{1}{N(T)+1}\sum_{j=0}^{N(T)}Y_j$. By Jensen's inequality, we have
\begin{align}
\mathbb{E}[Y_j^2]&=\frac{1}{N(T)+1}\sum_{j=0}^{N(T)}Y_j^2\geq (\mathbb{E}[Y_j])^2\nonumber\\
&=\frac{(\sum_{j=0}^{N(T)}Y_j)^2}{(N(T)+1)^2}=\frac{T^2}{(N(T)+1)^2}.\label{eq:Jeanse}
\end{align}

Equation \eqref{eq:Jeanse} gives us $\sum_{j=0}^{N(T)}Y_j^2\geq\frac{T^2}{N(T)+1}$, which can be substituted back to \eqref{eq:AoIboundder} as
\begin{align}
\bar{\Delta}_r&\geq \frac{1}{2}\lim_{T\rightarrow\infty}\frac{T}{N(T)+1}+\frac{1}{2}=\frac{1}{2}\lim_{T\rightarrow\infty}\frac{T}{N(T)}+\frac{1}{2}.\label{eq:lastStep}
\end{align}

Note that the term $\lim_{T\rightarrow\infty}\frac{N(T)}{T}$ is the long-term throughput of the network, which is bounded by the packet arrival rate and the capacity of the network. Specifically, we have
\begin{align}
&\lim_{T\rightarrow\infty}\frac{N(T)}{T}\leq \lambda \label{eq:arrivalcon}\\
&\lim_{T\rightarrow\infty}\frac{N(T)}{T}\leq \eta_{\max}(1-\epsilon),\label{eq:capacitycon}
\end{align}
where $\eta_{\max}(1-\epsilon)$ in \eqref{eq:capacitycon} is the effective capacity of the network under the resource constraint. Substituting \eqref{eq:arrivalcon} and \eqref{eq:capacitycon} into \eqref{eq:lastStep} renders the lower bound as
\begin{align}
\bar{\Delta}_r\geq \frac{1}{2}\left(\frac{1}{\min\{\lambda,\eta_{\max}(1-\epsilon)\}}+1\right). \label{eq:finalLB}
\end{align}

%As $T\rightarrow\infty$, the effect of the last $\widetilde{Y}$ on the long-term average AoI can be neglected, and hence we consider the receiver AoI in the first $\sum_{j=1}^{N(T)}Y_j$ time slot.

\section{Single Threshold Packet Scheduling Policy}
In this section, we propose a low-complexity single threshold packet scheduling policy which achieves near optimal AoI at a given resource constraint, and allows us to derive the closed-form expressions for the AoI distribution, average AoI and transmission cost as functions of the threshold. Hence, the tradeoff between the receiver AoI and transmission cost can be easily obtained by varying the threshold.
\subsection{Single Threshold Policy}
First, note that the transmitter will naturally choose to remain silent if the buffer at the transmitter is empty, i.e., the latest arrived packet has been successfully delivered, which happens with probability $P_{\mathrm{empty}}$. For all the best-effort transmission policies, we have
\begin{align}
P_{\mathrm{empty}}=\frac{(1-\varepsilon)(1-\lambda)}{1-\varepsilon+\varepsilon\lambda}.\label{eq:Pempty}
\end{align}

If the buffer is not empty, the optimal action at the $i$th time slot is determined based on the state $s_i=(\Delta_t(i),\Delta_r(i-1))$. If the transmitter decides to send a packet, it will receive a reward of $(-\Delta_t(i)-1)$ with probability $(1-\varepsilon)$, and a reward of $(-\Delta_r(i-1)-1)$ with probability $\varepsilon$. On the other hand, if the transmitter choose to keep silent, it will receive a reward of $(-\Delta_r(i-1)-1)$. The expected reward gain by sending the packet is hence given by
\begin{align}
\bar{R}_{i|a_i=1}=(1-\varepsilon)(\Delta_r(i-1)-\Delta_t(i)), \label{eq:expectedReward}
\end{align}
which is obtained at the constant cost of $C_i=1$. From \eqref{eq:expectedReward}, the expected AoI improvement by sending a packet is proportional to $(\Delta_r(i-1)-\Delta_t(i))$. Therefore, we propose a low-complexity single threshold policy in Definition~\ref{def:proposed}, which maximizes the expected reward gain under a given cost constraint.
\begin{definition}[Single threshold policy]\label{def:proposed}
At the $i$th time slot, if the buffer is not empty, the transmitter chooses to transmit the packet if and only if $(\Delta_r(i-1)-\Delta_t(i))$ is above certain threshold $\delta$, where $\delta$ is a positive integer. Mathematically,
\begin{align}
a_i=\begin{cases}
1, & \textnormal{ if }\Delta_r(i-1)-\Delta_t(i)\geq\delta\\
0, & \textnormal{otherwise}
\end{cases} \label{thr}
\end{align}
\end{definition}

Compared to the double threshold policy proposed in \cite{ref9}, the single threshold policy can be easily analyzed and it allows us to investigate the tradeoff between AoI and transmission cost more conveniently. Furthermore, the transmitter only needs to make the transmission decision at the time slot when new packet arrives. Specifically, assume a new packet arrives at the $i$th time slot, we have {$\Delta_t(i)=0$}. If the transmission is not activated, i.e., $\Delta_r(i-1)<\delta$, there will be no transmission activated until the next packet arrival, since both $\Delta_t$ and $\Delta_r$ grow linearly with time, resulting no change in their difference. On the other hand, if the transmission is activated at the $i$th time slot, i.e., {$\Delta_r(i-1)\geq\delta$}, this packet will be repeated until it is successfully delivered or preempted by the newly arrival packet, since the value $(\Delta_r(i-1)-\Delta_t(i))$ will not change until the packet is successfully delivered.

Denote by $P_j$ the probability that the AoI observed at the end of a typical time slot is $j$, i.e., $P_j\triangleq\Pr(\Delta_r=j),j=1,2,...$. Further denote by $P_{\Delta}$ the probability that the receiver AoI is greater or equal to $\delta$, i.e.,
\begin{align}
P_{\Delta}=\sum_{j=\delta}^{\infty}P_j.\label{eq:Pdelta}
\end{align}
 Based on the above analysis, the event $\Delta_r=j$ occurs in the following two cases.
\begin{itemize}
\item{\emph{Case 1:} The previous packet arrival occurs at $j$ time slots before, the transmission is activated when that packet arrives, and that packet has been successfully delivered during the subsequent $j$ time slots. Case~1 occurs with probability
    \begin{align}
    \bar{P}_{\mathrm{c1}}=P_{\Delta}(1-\lambda)^{j-1}\lambda(1-\varepsilon^j)
    \end{align} }
\item{\emph{Case 2:} The previous packet arrival occurs at $m$ time slots before, where $1\leq m\leq j-1$, and the observed AoI is $(j-m)$ when that packet arrives. Furthermore, no transmission is activated when the previous packet arrives or the transmission is activated, but all the $m$ attempts are erased. Case~2 occurs with probability
    \begin{align}
    \bar{P}_{\mathrm{c2}}=&\sum_{m=j-\delta+1}^{j-1}P_{j-m}(1-\lambda)^{m-1}\lambda \nonumber \\  &+\sum_{m=1}^{j-\delta}P_{j-m}(1-\lambda)^{m-1}\lambda\varepsilon^m
    \end{align}
    }
\end{itemize}

By combining the above two independent cases, we have
\begin{align}
P_j=\bar{P}_{\mathrm{c1}}+\bar{P}_{\mathrm{c2}}. \label{eq:Pj}
\end{align}

From \eqref{eq:Pdelta}-\eqref{eq:Pj}, we can solve for the distribution of $\Delta_r$, as given in Lemma~\ref{lem:Pj}. The detailed derivation are omitted for brevity.

\begin{lemma}\label{lem:Pj}
Consider a point-to-point communication system with Bernoulli packet arrival at rate $\lambda$ and random packet erasure channel with erasure probability $\varepsilon$.  If the single threshold policy with threshold $\delta$ is adopted, the probability mass function (PMF) of the receiver AoI $\Delta_r$ is given by
\begin{align}
&P_j\triangleq\Pr(\Delta_r=j)= \nonumber \\
&\begin{cases}
\frac{1}{\beta}(1-(1-\lambda)^j \varepsilon^j), & j \leq \delta\\
\frac{\lambda \varepsilon^{j-\delta+1}-\lambda \varepsilon^{\delta}(1-\lambda)^j+(1-\varepsilon)(1-\lambda)^{\delta}\varepsilon^j-(1-\varepsilon)(1-\lambda)^{j-\delta+1}}{\beta (\varepsilon+\lambda-1)}, & j>\delta.
\end{cases} \label{pmf}
\end{align}
where $\beta\triangleq\delta+\frac{\varepsilon}{1-\varepsilon}+\frac{(1-\varepsilon)(1-\lambda)}{(1-\varepsilon+\lambda\varepsilon)\lambda}+\frac{(1-\lambda)^{\delta}\varepsilon^{\delta}}{1-\varepsilon+\lambda\varepsilon}$.
\end{lemma}

Given the distribution of the receiver AoI, we are now ready to derive the long-term average AoI and transmission cost. Specifically, the average AoI is a function of threshold $\delta$, given by
\begin{align}
\bar{\Delta}_r(\delta)=\sum_{j=1}^{\infty}jP_j. \label{eq:avgAoI}
\end{align}

Note that the transmission is activated when the buffer is not empty and the difference of data freshness between the receiver and the transmitter is greater than or equal to the threshold $\delta$. These two independent events occur with probability $(1-P_{\mathrm{empty}})$ and $P_{\Delta}$, respectively. Hence, the long-term average transmission cost is also a function of threshold $\delta$, given by
\begin{align}
\eta(\delta)&=(1-P_{\mathrm{empty}})P_{\Delta}=\frac{\lambda}{1-\varepsilon+\lambda\varepsilon}\sum_{j=\delta}^{\infty}P_j\label{eq:avgCost}
\end{align}

By substituting $P_j$ derived from Lemma~\ref{lem:Pj} into \eqref{eq:avgAoI} and \eqref{eq:avgCost}, we can obtain the closed-form expressions of the average AoI and transmission cost, as given in {Lemma}~\ref{lem:avgAoI}.
\begin{lemma}\label{lem:avgAoI}
Consider a point-to-point communication system with Bernoulli packet arrival at rate $\lambda$ and random packet erasure channel with erasure probability $\varepsilon$.  If the single threshold policy with threshold $\delta$ is adopted, the long-term average receiver AoI and transmission cost are given by
\begin{align}
&\bar{\Delta}_r(\delta)=\frac{1}{\beta}\bigg(\frac{\delta(\delta+1)}{2}-\frac{(1-\lambda)\varepsilon}{(1-\varepsilon+\lambda\varepsilon)^2}(1-X)\nonumber\\
    &+\frac{\lambda\varepsilon^2}{(1-\varepsilon)^2(\varepsilon+\lambda-1)}-\frac{(1-\varepsilon)(1-\lambda)^2}{(\varepsilon+\lambda-1)\lambda^2}	+\frac{X}{(1-\varepsilon)\lambda}\bigg) \nonumber\\
    &+\frac{\delta}{\beta}\left(\frac{\lambda\varepsilon+(1-\varepsilon)(1-\lambda)}{(1-\varepsilon)\lambda}+\frac{1}{1-\varepsilon+\lambda\varepsilon}X\right)\label{eq:AoIFinal}\\
&\eta(\delta)=\frac{1}{(1-\varepsilon)\beta}\label{eq:etaFinal}
\end{align}
where $\beta$ is defined in Lemma 1 and $X\triangleq\varepsilon^{\delta}(1-\lambda)^{\delta}$.
\end{lemma}

\subsection{Selection of the Threshold}
It is observed that the average AoI $\bar{\Delta}_r(\delta)$ in \eqref{eq:AoIFinal} is a monotonically increasing function with the threshold $\delta$, while the transmission cost  $\eta(\delta)$ in \eqref{eq:etaFinal} is a monotonically decreasing function with $\delta$. To achieve the minimum AoI subject to cost constraint, we can find the maximum value of $\delta$ that meets the cost constraint. Mathematically, we have
\begin{align}
\bar{\Delta}_r^*=\min_{\delta}\bar{\Delta}_r(\delta).
\end{align}
The optimal threshold $\delta^*$ is solved from
\begin{align}
\delta^*=\{\arg\max_{\delta\in\mathbb{N}}\eta(\delta)\leq\eta_{\max}\},\label{eq:optThre}
\end{align}
where $\mathbb{N}$ is the natural number set. The proposed single threshold policy with the optimal threshold $\delta^*$ in \eqref{eq:optThre} is referred as the ``\emph{deterministic} single threshold policy". 

Note that the integer constraint on the threshold may lead to under-use of the resource, i.e., $\eta(\delta^*)<\eta_{\max}$. In such case, the AoI performance of the proposed transmission policy can be improved by introducing a ``\emph{random} policy" where the transmitter uses the threshold $\delta^*$ with probability $q$ and uses the threshold $(\delta^*+1)$ with probability $(1-q)$. The probability $q$ is determined by solving
\begin{align}
q\cdot\eta(\delta^*)+(1-q)\cdot\eta(\delta^*+1)=\eta_{\max}
\end{align}
The combination of the deterministic and the random single threshold policy mentioned above can achieve arbitrary point on the tradeoff curve between the AoI and transmission cost, and hence provide a complete solution to the problem $\mathbf{P0}$.

Note that if there is no resource constraint, we have $\delta^*=0$ and the proposed scheme reduces to the conventional pLGFS with performance given in \eqref{eq:pLGFS}. Furthermore, if we increase the packet arrival rate to $\lambda=1$, the network degenerates to the one with generate-at-will arrival model. The AoI and transmission cost equation in \eqref{eq:AoIFinal} and \eqref{eq:etaFinal} reduces to \eqref{eq:AoIARQ} and \eqref{eq:etaARQ}, respectively, which are proven to be optimal in \cite{ref14}.

\section{Numerical Results}
In this section, numerical results are presented to verify the analytical results and validate the performance of the proposed single threshold packet scheduling policy.

\begin{figure}[htbp]
	\centering
	\scalebox{0.8}{\includegraphics{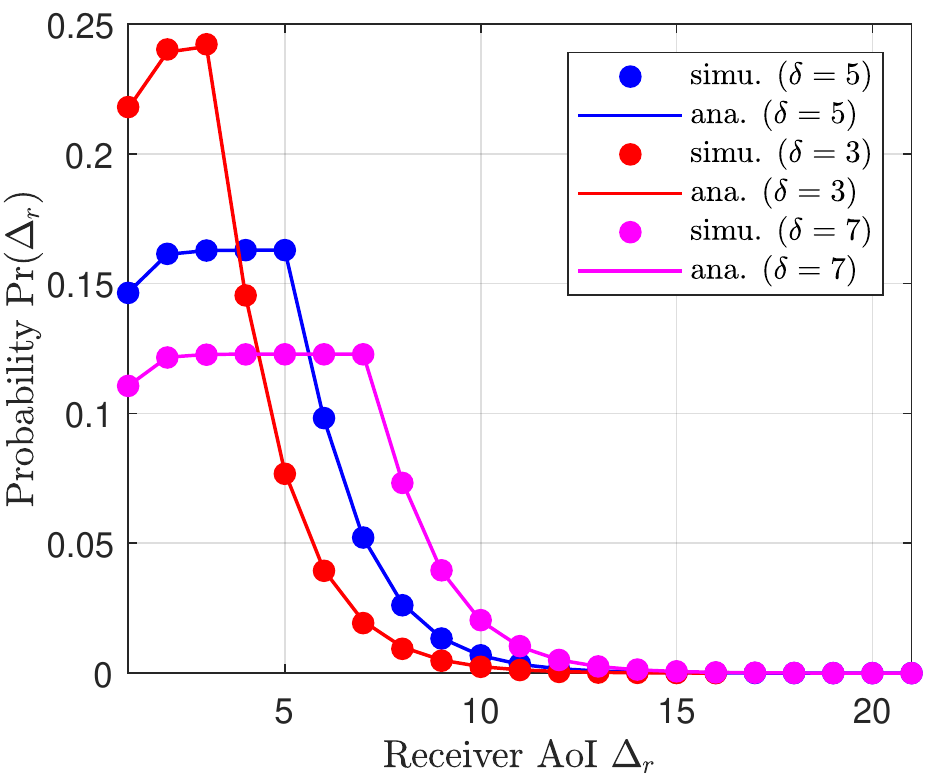}}
	\caption{Verify the analytical PMF of AoI with the network parameters set as $\lambda=0.5$ and $\varepsilon=0.2$.}
    \label{F:PMF}
\end{figure}

\begin{figure}[htbp]
	\centering
	\scalebox{0.6}{\includegraphics{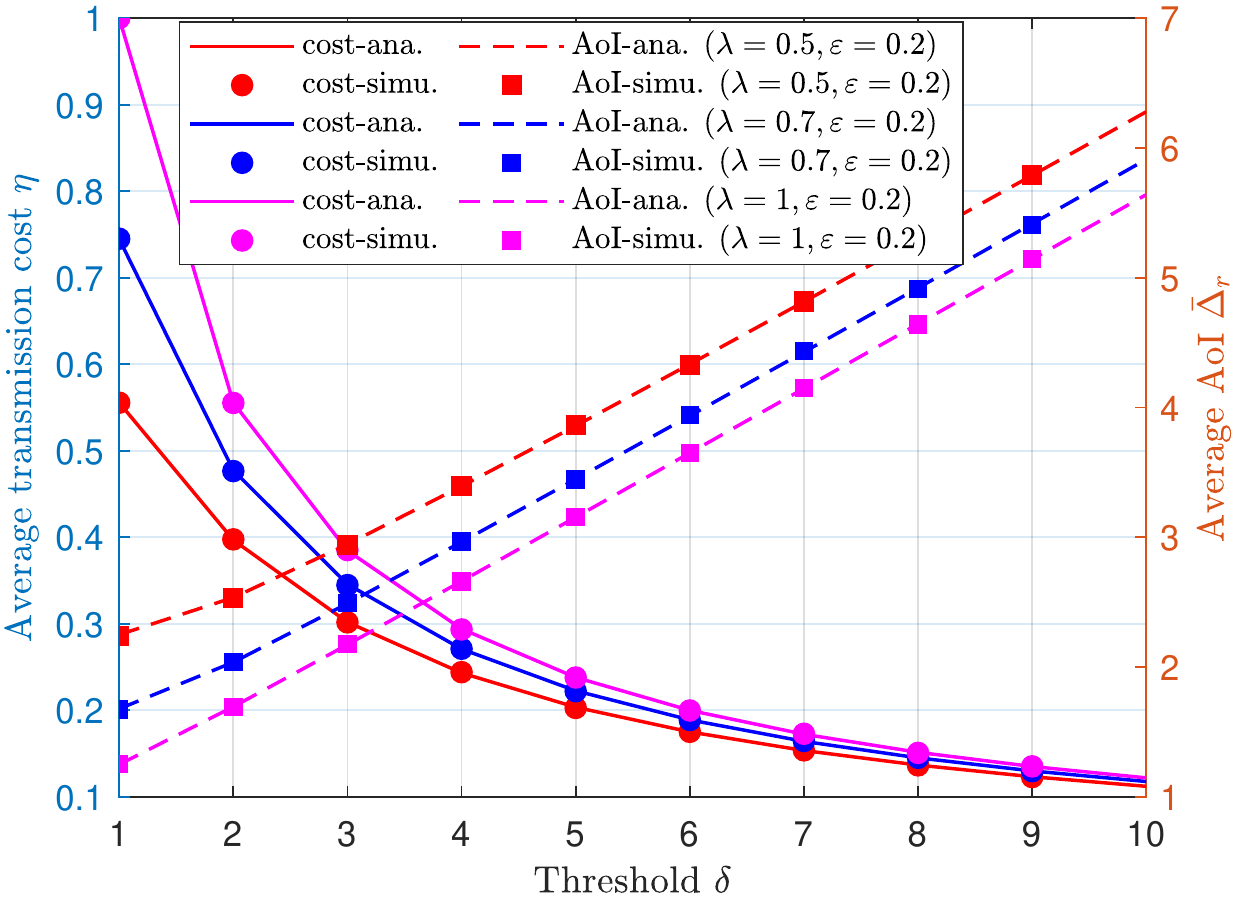}}
	\caption{Verify the analytical average AoI and transmission cost.}
    \label{F:avgAoI}
\end{figure}

First, we verify the analytical PMF of the AoI presented in Lemma~\ref{lem:Pj} with the simulation results in Fig.~\ref{F:PMF}. The simulation results are obtained with the observation interval $T=10^5$ time slots. It is observed that the analytical results match very well with the simulation results and the AoI tends to be larger with the increase of the threshold $\delta$. Next, the long-term average AoI and transmission cost presented in Lemma~\ref{lem:avgAoI} are verified with the simulation results in Fig.~\ref{F:avgAoI}. As expected, the average AoI monotonically increases with the threshold, while the transmission cost  monotonically decreases with the threshold.

\begin{figure}[htbp]
	\centering
	\scalebox{0.8}{\includegraphics{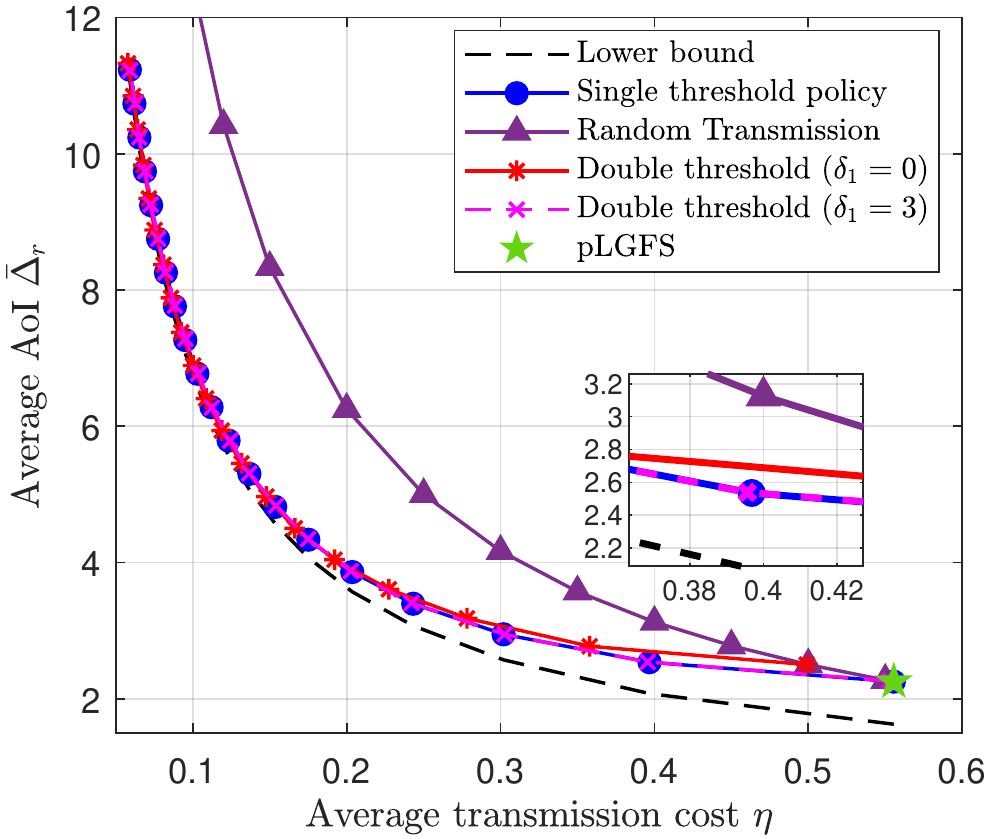}}
	\caption{Tradeoff between AoI and transmission cost achieved by various packet scheduling policies ($\lambda=0.5$, $\varepsilon=0.2$).}
    \label{F:compare}
\end{figure}

To validate the performance of the proposed single threshold packet scheduling policy, we compare the achievable tradeoff between the AoI and transmission cost with the lower bound in \eqref{eq:finalLB} and that achieved by the following benchmark schemes:
\begin{itemize}
\item{\emph{Random Transmission}: When the buffer is not empty, the transmitter chooses to activate the transmission with probability $\gamma$. Random transmission can be viewed as pLGFS with effective erasure probability $\varepsilon'=1-\gamma(1-\varepsilon)$. The average AoI and transmission cost are given by
    \begin{align}
    \bar{\Delta}_r'=\frac{1}{\lambda}+\frac{\varepsilon'}{1-\varepsilon'}, \textnormal{ and }\eta'=\frac{\gamma\lambda}{1-(1-\lambda)\varepsilon'}.
    \end{align}
    To ensure that the resource constraint is met, the transmission probability $\gamma$ is set by solving $\eta'=\eta_{\max}$, which renders
    \begin{align*}
    \gamma=\frac{\eta_{\max}\lambda}{\lambda-\eta_{\max}(1-\lambda)(1-\varepsilon)}.
    \end{align*}
     One of the advantages of the random transmission policy is that it does not rely on the feedback information.
     }

\item{\emph{Double Threshold Policy} \cite{ref9}: As suggested in \cite{ref9}, a double threshold method may further enhance the AoI performance. With double threshold policy, the transmitter chooses to send the packet in the buffer if the age of that packet is below certain threshold and the difference of data freshness between the transmitter and the receiver is above certain threshold, i.e., 
    \begin{align*}
    a_i=\begin{cases}1, & \textnormal{if }\Delta_t(i){\leq}\delta_1\ \&\ \Delta_r(i-1)-\Delta_t(i)\geq\delta_2\\
    0, &\textnormal{otherwise}
    \end{cases}
    \end{align*}
    Since \cite{ref9} considered the generate-at-will arrival model, the analytical results cannot be directly applied to the case with stochastic packet arrival. Hence, we choose the optimal threshold $(\delta_1,\delta_2)$ by exhaustive search in Fig.~\ref{F:compare}.}
\end{itemize}

It is observed from Fig.~\ref{F:compare} that the random transmission policy has the poorest performance, i.e., it has the largest AoI at the same amount of transmission cost. This is consistent with expectation since random transmission policy has neglected the environment status. The performance of the double threshold method is dependent on the selection of the thresholds. When the resource constraint is tight, it is desired to choose a small $\delta_1$. On the other hand, a larger value of $\delta_1$ is preferred when the resource constraint is loose. As compared with the double threshold policy, the proposed policy offers the benefit of simple implementation and closed-form performance expressions, which allow us to select the optimal threshold easily. It is observed from Fig.~\ref{F:compare} that the proposed single threshold policy has very close performance with the double threshold policy when $\delta_1=3$ and exhaustive search of $\delta_2$, and outperforms the double threshold policy when $\delta_1=0$.   The performance of the proposed single threshold policy is also very close to the lower bound, which further validates its optimality. Finally, note that all the schemes converge to the pLGFS when the maximum transmission cost is large enough, which is optimal with the absence of resource constraint.

\section{Conclusion}
This paper investigates the tradeoff between the long-term average AoI and transmission cost for a network with stochastic packet arrival and random erasure channel. We formulate the AoI minimization problem under the resource constraint as a CMDP and propose a low-complexity single threshold packet scheduling policy. The achievable AoI and transmission cost of the proposed scheme are expressed in simple closed-form. Simulation results show that the proposed single threshold policy has very close performance with the lower bound and the complex double threshold method with the optimal threshold pairs obtained via exhaustive search. Furthermore, the proposed policy degenerates to the optimal policies proposed in the literature under special cases.

Note that most of the packet scheduling policies discussed in this paper rely on the instantaneous feedback from the receiver, which may not be available in practice. In the future, we will extend the proposed scheme to the network where the feedback information is delayed or even absent, and investigate the tradeoff among more network performance metrics, e.g., the tradeoff between the receiver AoI and the long-term network throughput under resource constraint.
\section*{Acknowledgement}
This work was supported by the Natural Science Foundation of China with grant number 6504009712 and the Natural Science Foundation of Jiangsu Province with grant number BK20200354.

\bibliographystyle{ieeetr}
\bibliography{THE}

\end{document}